\documentclass{article}
\usepackage{spconf,amsmath,amssymb,graphicx,bm,mathtools,url,algorithm,algorithmic,xcolor}
\usepackage{flushend}

\usepackage{enumitem}
\setlist{nosep, leftmargin=14pt}

\begin{document}
\title{Maximizing Unambiguous Velocity Range in Phase-contrast MRI with Multipoint Encoding}

\name{Shen Zhao$^{\star}$ \qquad Rizwan Ahmad$^{\star \dagger}$ \qquad Lee C. Potter$^{\star}$ \thanks{This work was funded by the National Institute of Health grants R01HL135489 and R01HL151697.}}

\address{\emph{Address of author}}
\address{$^{\star}$The Ohio State University, Department of Electrical and Computer Engineering \\
 $^{\dagger}$The Ohio State University, Department of Biomedical Engineering}
\maketitle

\begin{abstract}
In phase-contrast magnetic resonance imaging (PC-MRI), the velocity of spins at a voxel is encoded in the image phase. The strength of the velocity encoding gradient offers a trade-off between the velocity-to-noise ratio (VNR) and the extent of phase aliasing. Phase differences provide invariance to an unknown background phase. Existing literature proposes processing a reduced set of phase difference equations, simplifying the phase unwrapping problem at the expense of VNR or unaliased range of velocities, or both.  Here, we demonstrate that the fullest unambiguous range of velocities is a parallelepiped, which can be accessed by jointly processing all phase differences.  The joint processing also maximizes the velocity-to-noise ratio.  The simple understanding of the unambiguous parallelepiped provides the potential for analyzing new multi-point acquisitions for an enhanced range of unaliased velocities; two examples are given.
\end{abstract}


\section{Introduction}
Phase-contrast magnetic resonance imaging is a non-invasive, quantitative technique to measure hemodynamics in vivo \cite{pelc1991phase,MarklSurvey,rich2016ab}. Velocity encoding is achieved via a time-varying magnetic field gradient, resulting in a per voxel phase that is related to the velocity of spins. Due to the existence of the background phase, linear proportionality between velocity and phase is achieved via conjugate multiplication of encodings. Thus, at least a 4-point encoding is necessary for 3-directional velocity mapping. For 4-point and 5-point encodings proposed in the literature, the set of many phase differences is pre-processed to yield fewer, simplified, equations in the three unknown velocity components.  

Building on our previous work \cite{zhao2021venc} for multi-point 1-directional velocity encoding, this paper demonstrates for multi-point 3-directional velocity encoding how all phase differences can be processed jointly to not only maximize velocity-to-noise ratio (VNR) but also maintain the fullest extent of the unambiguous range intrinsically provided by the encoding. Pre-processing typically reduces both VNR and the volume of unaliased velocities. Invoking results dating to Gauss 
\cite{Gauss1801}, the intrinsic unambiguous range is seen to be a parallelepiped. Moreover, we demonstrate how the set of unaliased velocities can be tremendously increased for multi-point encodings compared to past three decades' literature.

\section{Theory}
\subsection{Signal model}
A time-varying field gradient may be used to encode spin velocity into image phase. For $L$ encodings with first moments of time varying field gradient $\bm{m}_{l}$, and true velocity $\bm{v}$,  the complex-valued measurements at each spatial location are
\begin{equation}
    \label{eq: datamodel}
    \widetilde{y}_l = a_l e^{i (\phi_0 + \gamma \bm{m}_{l}^\intercal \bm{v} )} + n_l, \qquad l=1, ..., L, 
\end{equation}
where $a_l \in \mathbb{R}$ is noiseless signal magnitude, $\gamma$ is the gyromagnetic ratio, and $^{\intercal}$ denotes transpose. Here, $n_l \in \mathbb{C}$ is i.i.d. additive circularly symmetric complex Gaussian noise. 

\subsection{Congruence equations}
Without requiring additional measurement, to obtain equations only related to $\bm{v}$, the background phase $\phi_0$ is canceled via conjugate multiplication of points, $\widetilde{y}_l$. For $L$-point encoding, there are $N = \frac{L(L-1)}{2}$ pairs of phase differences, define
\begin{eqnarray}
    \bm{m}_{i,j}            &=& \bm{m}_i-\bm{m}_j \nonumber\\
    \bm{A}                  &=& \gamma\begin{bmatrix*}[r]
    \bm{m}_{2,1}            & \bm{m}_{3,1} & \cdots & \bm{m}_{L,L-1}
    \end{bmatrix*}^\intercal\nonumber \\
    \widetilde{\phi}_{i,j}  &=& \angle \widetilde{y}_i \widetilde{y}_j^* \nonumber\\
    \widetilde{\bm{\phi}}   &=& \begin{bmatrix*}[r] \widetilde{\phi}_{2,1} & \widetilde{\phi}_{3,1} & \cdots & \widetilde{\phi}_{L,L-1} \end{bmatrix*}^\intercal ,
\end{eqnarray}
where $(\cdot)^*$ denotes conjugation. We formulate the $N$ congruence equations
\begin{equation}
    \label{eqn: congruence}
    \bm{A} \bm{v} + \bm{\epsilon} \equiv \widetilde{\bm{\phi}} \mod 2\pi 
\end{equation}
where $\bm{\epsilon}$ is the noise, now correlated owing to the conjugate multiplies, and close to Gaussian for $\bm{n}$ is small. Considering the wrapping of phases, we can reformulate \eqref{eqn: congruence} as
\begin{eqnarray}
    \label{eqn: least square problem}
    \bm{A} \bm{v} + \bm{\epsilon}= \widetilde{\bm{\phi}} + 2\pi \bm{k} ,
\end{eqnarray}
where  $\bm{k}$ is a vector of wrapping integers. For multi-coil case, $\angle \widetilde{y}_i\widetilde{y}_j^*$ is replaced with the the angle of summation of conjugate multiplication across coils, detailed derivation is in \cite{zhao2021venc}.

\subsection{Joint solution}
For \eqref{eqn: least square problem}, the solution technique in \cite{zhao2021venc} may be extended to provide an approximate 
maximum likelihood estimate of the three velocity components $\bm{v}$. The solution is found 
by searching a small set of possible wrapping integers and computing a weighted linear combination of the $N$ phase differences:
\begin{equation}
\label{eqn:amle}
    \widehat{\bm{v}}  = \left( \bm{A}^{\intercal} \bm{\Sigma}^{-1} \bm{A} \right)^{-1} \bm{A}^{\intercal}  \bm{\Sigma}^{-1} \left( \widetilde{\bm{\phi}} + 2 \pi  \widehat{\bm{k}} \right).
\end{equation}
The estimator incorporates correlation matrix, $\bm{\Sigma} \in \mathbb{R}^{N\times N}$, of the noise $\bm{\epsilon}$. Assuming correct detection of the wrapping integers, $\bm{k}$, the noise sensitivity, $\eta$, reported as the volume of an uncertainty ellipsoid around the velocity estimate, is 
\begin{equation}
    \label{eqn:noiseball}
    \eta = \left( \det  \bm{A}^{\intercal} \bm{\Sigma}^{-1} \bm{A} \right)^{-1}.
\end{equation}
Let $\bm{P}$ denote any $D$-by-$N$ matrix
that pre-processes the $N$ phase differences into $3 \leq D\leq N$ new equations,  
$ \bm{P} \bm{A} \bm{v} +\bm{P}\bm{\epsilon} \equiv \bm{P} \widetilde{\bm{\phi}} 
+ 2\pi \bm{P} \bm{k}$. The noise sensitivity, $\eta_{\bm{P}}$, then becomes
\begin{equation}
    \label{eqn:noiseball2}
    \eta_{\bm{P}} = \left( \det \bm{A}^\intercal \bm{P}^\intercal \left(\bm{P \Sigma} \bm{P}^\intercal \right)^{-1} \bm{PA} \right)^{-1}. 
\end{equation}
Note from the data processing inequality \cite{CoverIE} that $\eta_{\bm{P}} \geq \eta$ for any $\bm{P}$.  Additionally, assessment of \ref{eqn:amle} for candidate wrapping integers also yields the probabilities of wrapping errors, given the noise power in the complex-valued data, $\widetilde{y}_l$.

\subsection{Unambiguous range}
\label{sec:lattice}
We assume that the matrix $\bm{A}$ has rank 3, so that no direction is invisible to the data acquisition. For the noiseless congruence problem $\bm{Av} \equiv \bm{0} \mod 2\pi$, all solutions $\bm{v}$ in 3D Euclidean space, $\mathbb{R}^3$, make a periodic lattice $\bm{\Lambda}$~\cite{Gauss1801,Barvinok, knill2012multivariable}. Given phase differences $\widetilde{\bm{\phi}}$ and a corresponding solution $\bm{v}_\star$ to \eqref{eqn: congruence}, addition of any lattice point to $\bm{v}_\star$ yields another solution; thus, the congruence equations are ambiguous for the velocity. There exist three vectors $\bm{v}_1,\bm{v}_2,\bm{v}_3$ such that all points $\bm{v} \in \bm{\bm{\Lambda}}$ admit a unique representation
\begin{eqnarray}
    \bm{v} = \sum_{i=1}^3 p_i \bm{v}_i,~ p_i \in \mathbb{Z},
\end{eqnarray}
where $\mathbb{Z}$ is the set of integers. The triple of vectors $\bm{V} = [\bm{v}_1,\bm{v}_2,\bm{v}_3]$ makes a $3$-by-$3$ matrix and is called a basis of $\bm{\Lambda}$. Following \cite{Barvinok}, we define the set
\begin{eqnarray}
    \Omega = \left\{\sum_{i=1}^3 \alpha_i \bm{v}_i,~ \alpha_i \in [0,1) \right\}
\end{eqnarray}
as a \emph{fundamental parallelepiped} of the 
lattice $\bm{\Lambda}$. This implies that the shifts of $\Omega$, $\{\Omega+\bm{v}, \bm{v}\in \bm{\Lambda}\}$, cover $\mathbb{R}^3$ without overlapping. We call the set $\Omega$ an \emph{unambiguous range} for the congruence equations specified by $\bm{A}$.

\begin{figure}[htb]
\begin{center}
    \includegraphics[width=\columnwidth]{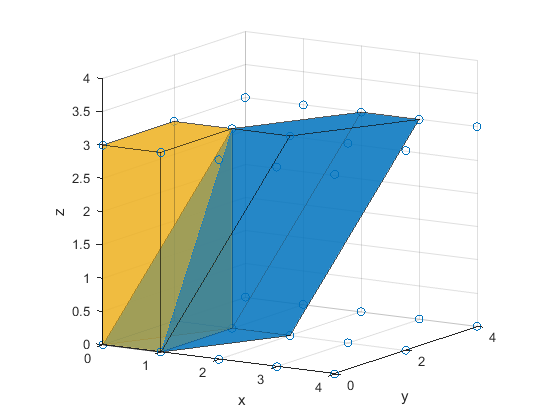}
    \caption{Illustration of two possible fundamental parallelepipeds defining an unambiguous range for $\bm{A}$ given in \eqref{eqn: example2}.  The lattice $\bm{\Lambda}$ is denoted by blue circles.
    The volume of each 3D fundamental parallelepiped is the same. }
    \label{fig: possible parallelepiped}
\end{center}
\end{figure}

The choice of basis $\bm{V}$ and corresponding parallelepiped is not unique; but, all choices have the same volume, denoted $|\Omega|$ and equal to $|\det(\bm{V})|$. Among all possible $\bm{V}$, we choose the $\bm{V}$ with the smallest condition number, i.e., we choose $\Omega$ closest to a cube to spread the unambiguous velocity coverage in a nearly isotropic manner. However, choices of a parallelepiped with a long and ``pointy'' shape along a certain direction might be desirable for a specific application. Examples of two
possible fundamental parallelepipeds are shown in Fig.~\ref{fig: possible parallelepiped}
for the encoding matrix
\begin{equation}
\label{eqn: example2}
\bm{A} = 2\pi \begin{bmatrix*}[r] 1 & 0 & 0\\ 0 & \frac{1}{2} & 0\\ 0 & 0 & \frac{1}{3} \end{bmatrix*}.
\end{equation}

To numerically construct a parallelepiped for the unambiguous range, we set $\widetilde{\bm{\phi}} = \bm{0}$ and search over $\bm{\Lambda}$. The search can be simplified by finding an upper bound on $\Omega$. To handle the case that an element of $\bm{A}$ may be zero, we define $\oslash$ to be the modified element-wise division, and $\overline{\text{lcm}}$ as a generalized least common multiple (lcm) of numbers that can include $0$: 
\begin{eqnarray}
    \bm{c} &=& a \oslash \bm{b}, \text{where } c_i = 
    \begin{cases} \frac{a}{b_i}, \text{ if } b_i \not = 0 \\
    0, \text{ else}
    \end{cases}\nonumber\\
    \overline{\text{lcm}} (\{b_i\}) &=&  \text{lcm} (\{b_i|b_i \not = 0\}) .
\end{eqnarray}
For convenience, we use $\bm{A}_{:,i}$ and $ \bm{A}_{i,:}$ to denote the $i^{th}$ column and row of $\bm{A}$, respectively. Thus, we can observe that if $\bm{A}\bm{v}_\star \equiv \bm{0} \mod 2\pi$, then
\begin{eqnarray}
    \bm{A} 
    \left(\bm{v}_\star + 
    \begin{bmatrix*}[r] 
    ~\overline{\text{lcm}}(2\pi \oslash \bm{A}_{:,1}) \\ 
    ~\overline{\text{lcm}}(2\pi \oslash \bm{A}_{:,2}) \\
    ~\overline{\text{lcm}}(2\pi \oslash \bm{A}_{:,3}) 
    \end{bmatrix*}
    \right)
    \equiv \bm{0} \mod 2\pi .
\end{eqnarray}
Let $\Delta$ be a hyper-rectangle in $\mathbb{R}^3$ with edge along dimension $i$ given by
\begin{eqnarray}
    \left[-\overline{\text{lcm}}(2\pi \oslash \bm{A}_{:,i}), ~\overline{\text{lcm}}(2\pi \oslash \bm{A}_{:,i})\right] .
\end{eqnarray}
Observe that basis vectors $\bm{v}_1, \bm{v}_2,\bm{v}_3 \in \Omega \subseteq \Delta$ are solutions to $\bm{A}\bm{v} \equiv \bm{0} \mod 2\pi$. Then we can numerically find solutions of \eqref{eqn: least square problem} inside $\Delta$ by searching all possible choices of wrapping integers $\bm{k}$ determined by $\left\lfloor \frac{1}{2\pi} \bm{A}\Delta \right\rfloor$, where $\lfloor \cdot \rfloor$ is floor function. Among solutions achieving the minimum volume, we choose one with the smallest condition number, $\kappa (\bm{V})$.

\section{Methods and Results}
We review the pre-processed congruence equations formulated in existing methods and compare
the unambiguous ranges to $\Omega$ achieved by jointly solving all $N$ congruence equations
in \eqref{eqn: congruence}.
Additionally, we illustrate how the fundamental parallelepiped can be used to characterize $\Omega$
for any modified acquisition, potentially providing a significant increase in the unambiguous range.

\subsection{Balanced 4-point encoding}
Balanced 4-point encoding uses vertexes of a regular tetrahedron:
\begin{eqnarray}
    \bm{M} = m\begin{bmatrix*}[r]
    -1  & -1    & -1\\
    +1  & +1    & -1\\
    +1  & -1    & +1\\
    -1  & +1    & +1
    \end{bmatrix*},
    \bm{A} =\gamma m\begin{bmatrix*}[r]
    2   &2  &0\\
    2   &0  &2\\
    0   &2  &2\\
    0   &-2 &2\\
    -2  &0  &2\\
    -2  &2  &0
    \end{bmatrix*}
\end{eqnarray}
where $m$ is a positive constant. In 1991, Pelc et al.~\cite{pelc1991encoding} introduced 3-directional flow encoding and pre-processed the phase differences to conveniently yield a decoupled set of three equations in three unknown velocity components. Specifically, the pre-processing is given by $\bm{P}_{91} \in \mathbb{R}^{3\times 6}$:
\begin{eqnarray}
    \bm{P}_{91} \bm{A} \bm{v} +\bm{P}_{91}\bm{\epsilon} \equiv \bm{P}_{91} \widetilde{\bm{\phi}} \mod 2\pi&\\
    \bm{P}_{91} = \begin{bmatrix*}[r]
        1   &0  &0  &0  &0  &-1\\
        1   &0  &0  &0  &0  &1 \\
        0   &0  &1  &1  &0  &0
    \end{bmatrix*}&\\
    \label{eqn: pelc P}
    \bm{P}_{91}\bm{A} = \gamma m\begin{bmatrix*}[r]
    4   &0  &0\\
    0   &4  &0\\
    0   &0  &4
    \end{bmatrix*}&.
\end{eqnarray}
The pre-processing diminishes the information content. Starting from \eqref{eqn: pelc P}, the noise sensitivity is  degraded by $3\%$ compared to using all six phase differences, for a signal-to-noise ratio (SNR) of
$\sqrt{a_l^2/\text{var}(n_l)}=5$. The unambiguous set of velocities is reduced to a cube of edge length $\pi (2\gamma m)^{-1}$. 

In 2010, Johnson and Markl \cite{johnson2010improved} proposed a pre-processing that keeps the first three rows of $\bm{A}$, resulting in three coupled equations in three unknowns: 
\begin{eqnarray}
    \bm{P}_{10} \bm{A} v +\bm{P}_{10}\bm{\epsilon} \equiv \bm{P}_{10} \widetilde{\phi} \mod 2\pi&\\
    \bm{P}_{10} = \begin{bmatrix*}[r]
        1   &0  &0  &0  &0  &0\\
        0   &1  &0  &0  &0  &0 \\
        0   &0  &1  &0  &0  &0
    \end{bmatrix*}&\\
    \bm{P}_{10}\bm{A} = \gamma m\begin{bmatrix*}[r]
    2   &2  &0\\
    2   &0  &2\\
    0   &2  &2
    \end{bmatrix*}& .
\end{eqnarray}
To conveniently eliminate the need for phase unwrapping, the unambiguous set of velocities is defined in \cite{johnson2010improved} such that $-\pi\preccurlyeq \bm{P}_{10} \bm{A} \bm{v} \preccurlyeq \pi$
\begin{equation}
    \label{eqn:JM}
    -\pi\preccurlyeq \bm{P}_{10} \bm{A} \bm{v} \preccurlyeq \pi ,
\end{equation}
where $\preccurlyeq$ is element-wise inequality.  Thus, the unambiguous set in \cite{johnson2010improved} is determined by six linear inequalities.  The pre-processing slightly worsens noise sensitivity by $6\%$ at SNR of 5,compared to using all six phase differences; interestingly, the pre-processing and restriction in \eqref{eqn:JM} in this case preserve the fundamental parallelepiped for the encoding matrix $\bm{A}$.

\begin{figure}[htb]
    \centering
    \includegraphics[width = \columnwidth]{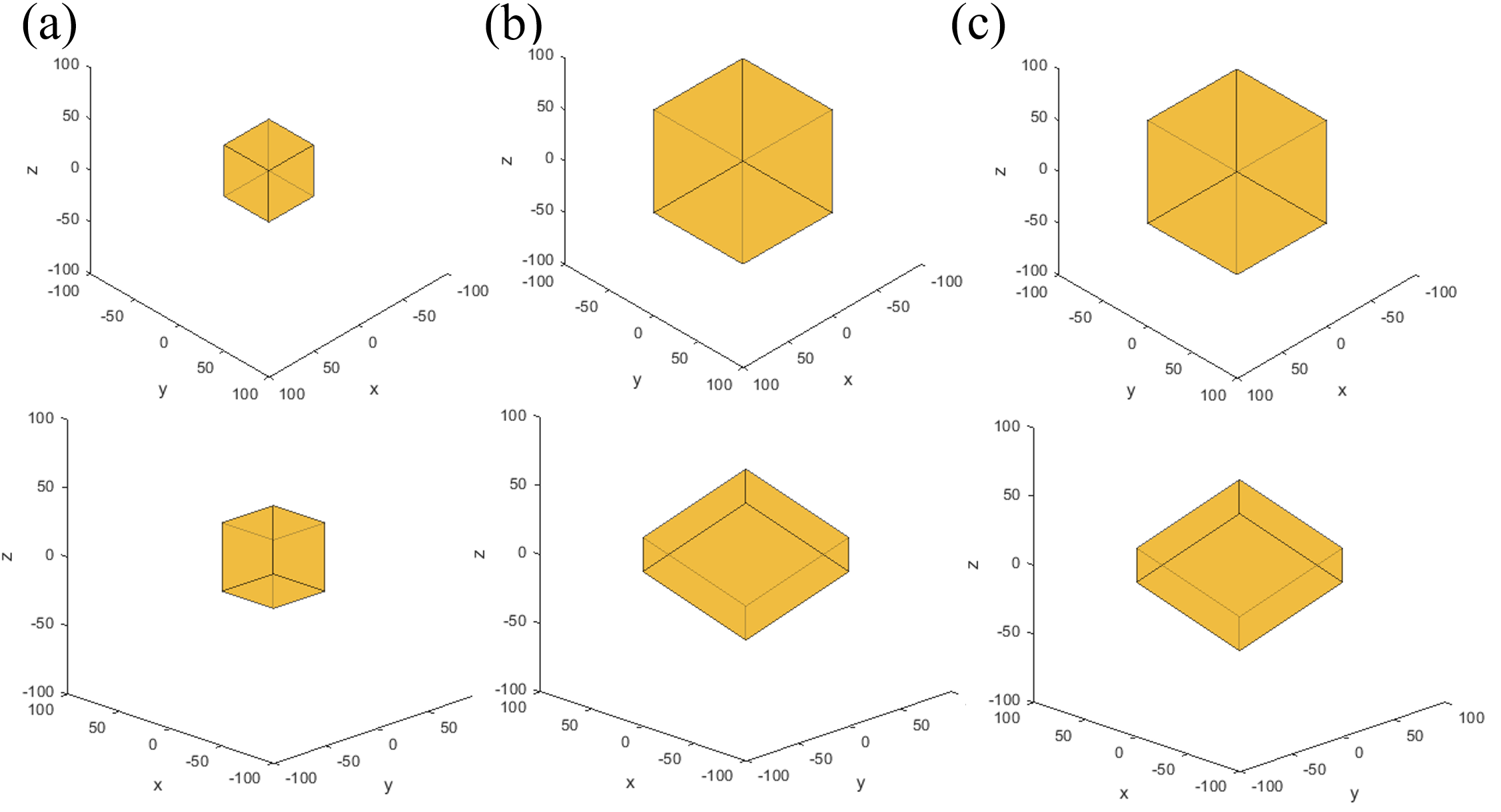}
    \caption{Balanced 4-point encoding with $\gamma m = \frac{\pi}{100}$. The rows employ two different camera lines of sight. The unambiguous velocity ranges in (a), (b), (c) are, respectively, for pre-processing $\bm{P}_{91}$, pre-processing $\bm{P}_{10}$, and using all six phase differences.}
    \label{fig: 4 point}
\end{figure}

In Fig.~\ref{fig: 4 point}, we illustrate the unambiguous velocity range for 4-point encoding and compare with the pre-processing approaches reviewed above. The pre-processing $\bm{P}_{91}$ leads to a 4 times smaller parallelepiped volume compared to both the fundamental parallelepiped, $\Omega$, and the pre-processing with $\bm{P}_{10}$.

\subsection{Balanced 5-point encoding}
\begin{figure}[tbh]
    \centering
    \includegraphics[width = \columnwidth]{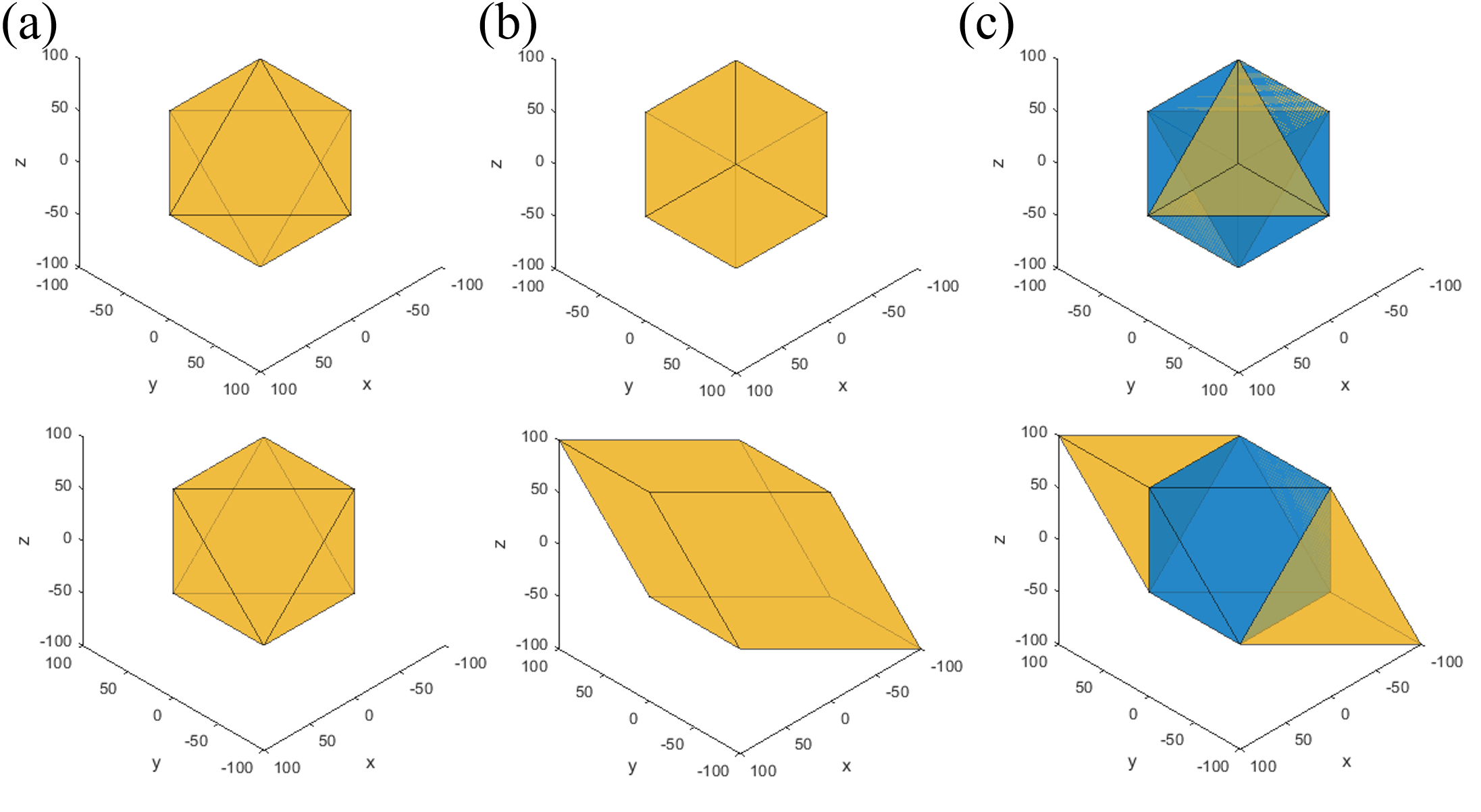}
    \caption{Balanced 5-point encoding with $\gamma m = \frac{\pi}{100}$. The first and the second rows employ two different camera lines of sight. The unambiguous velocity ranges in (a) and (b) are, respectively, for pre-processing $\bm{P}_{5}$ and use of all phase differences.
    Column (c) displays the volumes in (a) and (b) superimposed.}
    \label{fig: 5 point}
\end{figure}

Balanced 5-point encoding \cite{johnson2010improved} augments the balanced 4-point encoding with an additional point at the origin:
\begin{eqnarray}
    \bm{M} = m\begin{bmatrix*}[r]
    0   &  0    &  0\\
    -1  & -1    & -1\\
    +1  & +1    & -1\\
    +1  & -1    & +1\\
    -1  & +1    & +1
    \end{bmatrix*},
    \bm{A} = \gamma m
    \begin{small}
    \begin{bmatrix*}[r]
    -1      &-1     &-1\\
     1      &1      &-1\\
     1      &-1     &1\\
    -1      & 1     &1\\
     2      &2      &0\\
     2      &0      &2\\
     0      &2      &2\\
     0      &-2     &2\\
    -2      &0      &2\\
    -2      &2      &0
    \end{bmatrix*}
    \end{small}.
\end{eqnarray}
The pre-processing proposed in \cite{johnson2010improved} for 5-point encoding utilizes the first four equations of $\bm{A}$ to determine the unambiguous velocity range.
\begin{eqnarray}
    \bm{P}_{5} \bm{A} v +\bm{P}_{5}\bm{\epsilon} \equiv \bm{P}_{5} \widetilde{\bm{\phi}} \mod 2\pi&\\
    \bm{P}_{5} = \begin{bmatrix*}[r]
        \bm{I}_{4 \times 4} & \bm{0}_{4\times 6}
    \end{bmatrix*}&\\
    \bm{P}_{5}\bm{A} = \gamma m\begin{bmatrix*}[r]
        -1  & -1    & -1\\
        1   & 1     & -1\\
        1   & -1    & 1\\
        -1  & 1     & 1
    \end{bmatrix*}&
\end{eqnarray}
The four equations in three unknowns are solved least-squares, again with the restriction
\begin{equation}
    \label{eqn:JM2}
    -\pi\preccurlyeq \gamma\bm{P}_{5} \bm{A} \bm{v} \preccurlyeq \pi .
\end{equation}
In Fig.~\ref{fig: 5 point}, we illustrate 
the unambiguous velocity range for the $\bm{P}_{5}$  pre-processing with restriction in \eqref{eqn:JM2}. Using all phase differences leads to
$1.5$ times larger volume and $10\%$ better noise sensitivity, compared to the $\bm{P}_{5}$ pre-processing.

\subsection{Perturbed 4- and 5-point encodings}
We have shown that in phase-contrast-based flow imaging, all phase differences can be processed jointly to maintain the noise sensitivity and full unambiguous range intrinsic to the encoding matrix, $\bm{A}$.  In this subsection, we illustrate the potential of utilizing simple lattice concept in Section~\ref{sec:lattice} to enable encoding design resulting in an increased unambiguous velocity range, $\Omega$.

\begin{figure}[htb]
    \centering
    \includegraphics[width = \columnwidth]{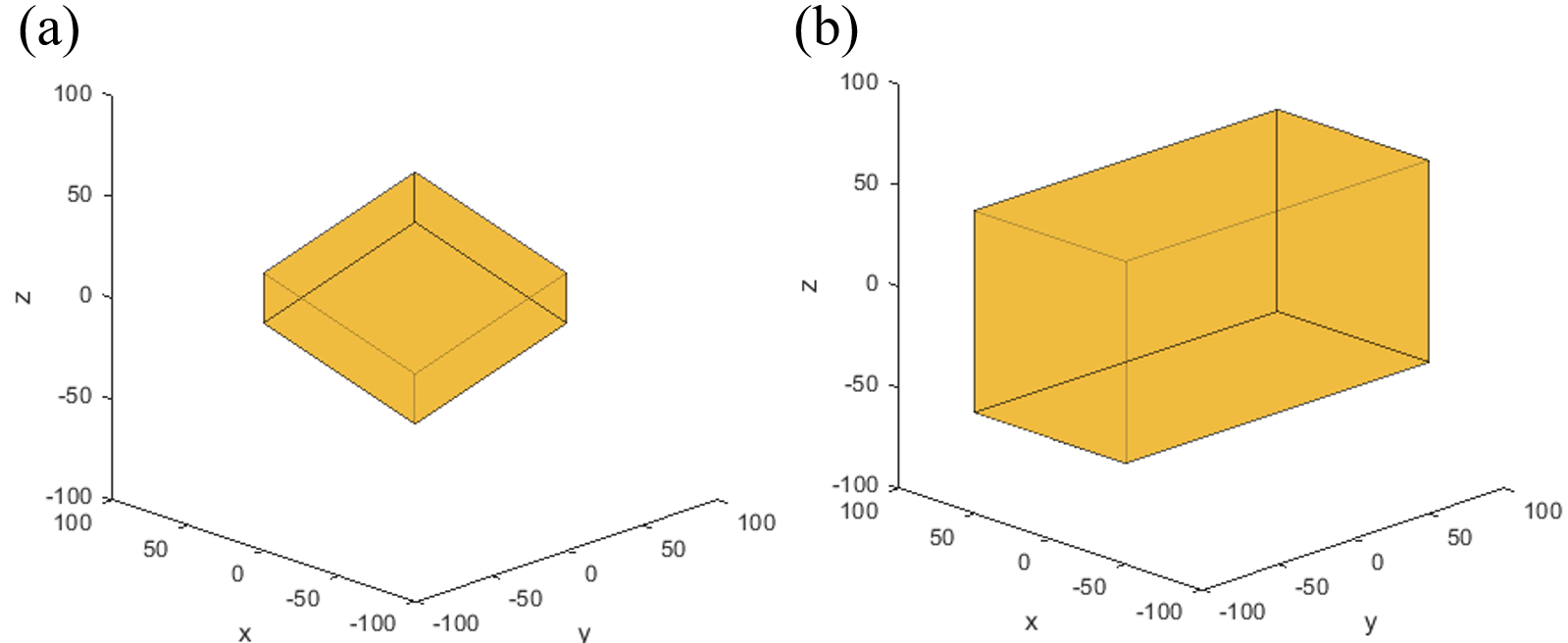}
    \caption{Unambiguous velocity ranges with $\gamma m = \frac{\pi}{100}$: (a) balanced 4-point encoding, $\bm{M}$; (b) perturbed 4-point encoding $\bm{M}'$. }
    \label{fig: 4 point perturbed}
\end{figure}

For 4-point encoding, we replace the first row of $\bm{M}$ with $\left[ -1, \, -0.5, \, -1\right]$ to obtain $\bm{M}'$. Then after processing all phase differences, the unambiguous range $\Omega$ for $\bm{M}'$ is $4$ times larger than for $\bm{M}$, as illustrated in Fig.~\ref{fig: 4 point perturbed}.

\begin{figure}[htb]
    \centering
    \includegraphics[width = 0.9\columnwidth]{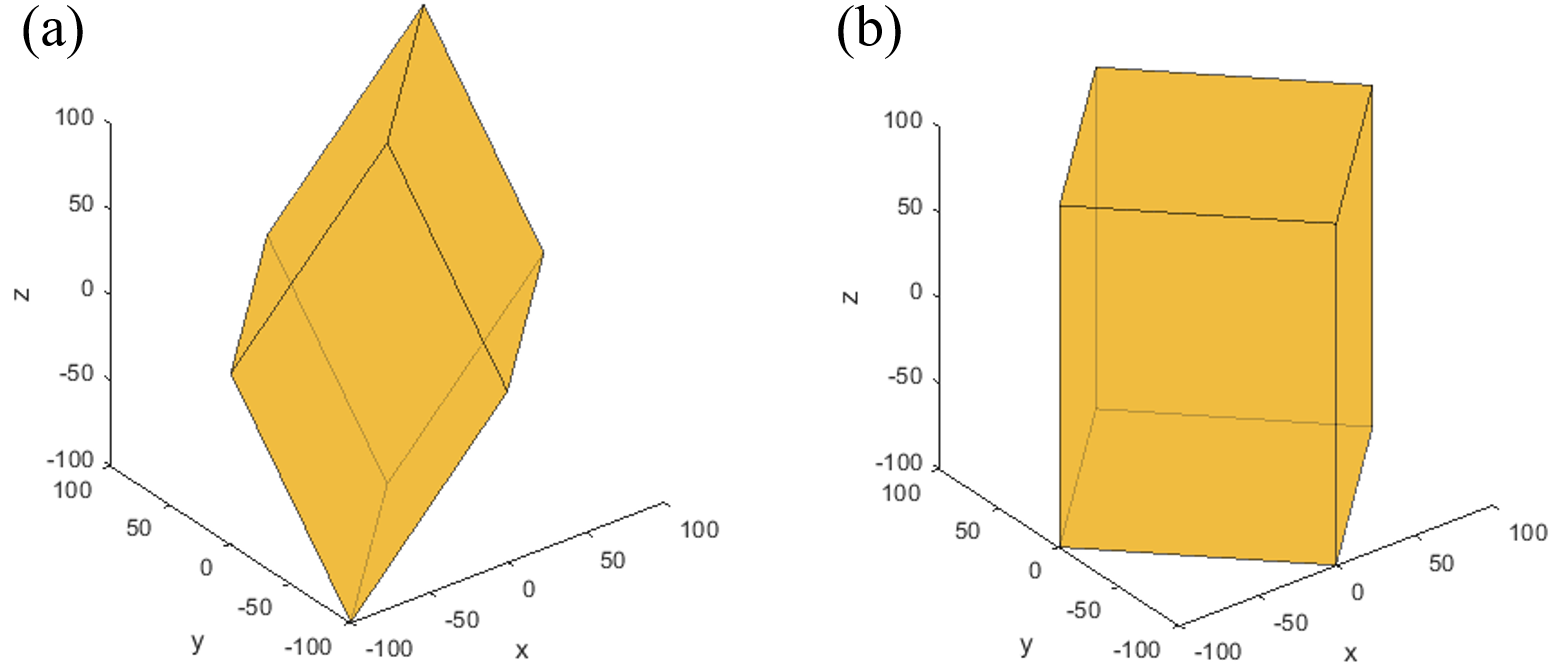}
       \caption{Unambiguous velocity ranges with $\gamma m = \frac{\pi}{100}$: (a) balanced 5-point encoding, $\bm{M}$; (b) perturbed 5-point encoding $\bm{M}'$. }
    \label{fig: 5 point perturbed}
\end{figure}

For 5-point encoding, we replace the first row of $\bm{M}$ with $\left[ 0, \, 0, \, 0.4\right]$ to obtain $\bm{M}'$. Then after processing all phase differences, 
the unambiguous range $\Omega$ for $\bm{M}'$ is $2$ times larger, as illustrated in Fig.~\ref{fig: 5 point perturbed}.

These two examples are suggestive of possible improvements; optimized acquisition would combine unambiguous range given by the parallelepiped $\Omega$, noise sensitivity in \eqref{eqn:noiseball}, and probability of wrapping errors \cite{zhao2021venc}. The optimized design is beyond the scope of this short conference manuscript. In conclusion, we have demonstrated that compared to the existing practice of employing a pre-processing step, jointly processing all phase differences leads to a larger unambiguous velocity range and higher VNR.

\clearpage
\section{Compliance with Ethical Standards}
This is a numerical simulation study for which no ethical approval was required.

\section{Acknowledgments}
\label{sec:acknowledgments}
The authors have no relevant financial or non-financial interests to
disclose.




\bibliographystyle{IEEEbib}
\bibliography{strings,refs}
\end{document}